\documentclass[12pt]{iopart}
\usepackage{amssymb}
\usepackage{amsfonts}
\usepackage{graphicx}

\font\frak=eufm10 scaled\magstep1
\def\goth#1{\hbox{{\frak#1}}}
\def\matriz#1#2{\left( \begin{array}{#1} #2 \end{array}\right) }
\def\GR{{\mathcal{G}}}
\def\dfrac#1#2{{\displaystyle\frac{#1}{#2}}}
\def\pd#1#2{\frac{\partial#1}{\partial#2}}

\newtheorem{theorem}{Theorem}

\begin{document}
\title{Integrability of Lie systems and some 
of its  applications in physics}
 \author{Jos\'e F. Cari\~nena, Javier de Lucas and Manuel F. Ra\~nada}

\address{Departamento de F\'{\i}sica Te\'orica, Facultad de Ciencias,
   Universidad de Zaragoza, 50009 Zaragoza, Spain}

\ead{jfc@unizar.es, dlucas@unizar.es, mfran@unizar.es}

\begin{abstract}
The geometric theory of Lie systems will be used to establish
 integrability conditions for several systems of differential equations, in
 particular,  Riccati
equations and Ermakov systems. Many different integrability criteria in the
literature will be analysed from this new perspective and some applications in physics
will be given.
\end{abstract}

\section{Introduction: Non-autonomous systems of differential equations}

Non-autonomous systems of differential equations play a relevant r\^ole in 
many physical theories, describing evolution  in terms of either time  or any
other parameter. For instance 
 Hamilton equations are 
systems of first-order differential equations and  Euler--Lagrange 
equations for regular Lagrangians are systems of second-order
differential  equations. These systems also appear in many other problems, and
the theorem of local
 existence and uniqueness of solutions is of a crucial importance. However, it
 is not possible to finding explicit solutions  of such systems and
 approximation methods have been developed for dealing with them. This gives
 even more relevance to the explicitly solvable examples which can be used to
 test approximation methods. In  the linear case there is a 
linear superposition principle allowing us to write the  general solution as a
linear combination of a fundamental set of solutions. Particular
solutions are also used to reduce the problem to a simpler one (see
e.g. \cite{CarRam} for Riccati equations). Our aim in this paper is to analyse 
sufficient conditions for integrability of a particular case of non autonomous
systems, the so called Lie systems \cite{LS,{Win83},{CGM00},{CGM07}}.
We shall understand that to find a solution means to reduce the problem  to some quadratures.

We remark that a  system of second-order differential equations in $n$  variables is related with a system of
first-order equations in $2n$ variables,
\begin{equation}
\ddot x^i=F^i(x,\dot x,t)\Longrightarrow\left\{\begin{array}{rcl} \dot
    x^i&=&v^i\\ \dot v^i&=&F^i(x,v,t)\end{array}\right. \,,\qquad i=1,\ldots n\,.\label{nonasodesys}
\end{equation}
and therefore it is enough to restrict ourselves to study systems of  first-order 
differential equations. For instance,  for the harmonic oscillator with a $t$-dependent angular frequency
$\omega(t)$:
\begin{equation}
\ddot x= -\omega^2(t)\, x
\Longleftrightarrow \left\{\begin{array}{rcl}\dot x&=&v\cr \dot
    v&=&-\omega^2(t)\, x\end{array}\right..\label{howt}
\end{equation}

 {}From the geometric viewpoint, a system
\begin{equation}
\dot x^i=X^i(x,t)\,,\qquad i=1,\ldots,n\,,\label{nonasys}
\end{equation}
is associated with the $t$-dependent vector field 
\begin{equation}
X= X^i(x,t)\,\pd{}{x^i}\label{tdepX}
\end{equation}
whose integral curves are determined by the solutions of (\ref{nonasys}). In
particular, for (\ref{howt}),
$$X=v\pd{}x -\omega^2(t) x\, \pd{}v\ ,
$$ 

The general solution of the inhomogeneous linear differential equation
$$
\frac{dx}{dt}=b_0(t)+b_1(t)x
$$
can be found with  two quadratures. It is 
given by 
$$
x(t)=\exp\left(\int_0^tb_1(s)\,ds\,\right) \times \left(x_0+\int_0^t b_0(t^{\prime})
        \exp\left(- \int_0^{t^{\prime}} b_1(s) \,ds \right) dt^{\prime}
\right)\ ,
$$
and  when the systems we are dealing with are linear, there is a {\sl
  linear
superposition principle}
allowing us to find the general solution as a linear combination of $n$
particular solutions.
For instance, for (\ref{howt}) if we know a particular solution the general solution can be found by
means of one quadrature, while if we know two particular solutions, $x_1$ and $x_2$, 
the general solution is a linear
combination (no quadrature is needed) 
$x(t)=k_1\,x_1(t)+k_2\, x_2(t)$. A less known fact is the
existence of other systems for which one can write the general solution as a
(maybe nonlinear) function of a fundamental set of solutions and a set of
constants determining each particular solution. The characterization of such
systems is due to Lie \cite{LS} and these Lie systems appear quite often not
only in mathematics but also in both
classical and quantum physics \cite{{CGM00},{CGM07},{AHW81},{HarWinAnd83},{SW84},{OlmRodWin86a},{OlmRodWin86b},{OlmRodWin87}}, the almost
 ubiquitous Riccati equation,
\begin{equation}
\frac{dx(t)}{dt}=b_2(t)\,x^2(t)+b_1(t)\,x(t)+ b_0(t)\ .
\label{ricceq}
\end{equation}
which appears in the reduction process from a linear second-order equation when
taking into account its dilation invariance, or in factorization methods,
being one important example \cite{{CarRam},{AHW81},{CarRamcinc},{CLR07b}}. 
 For such Riccati  equation, 
 if a particular solution is known, the general solution is obtained by two
quadratures; if two particular solutions are known the problem reduces to 
one quadrature; and finally,  when three particular solutions 
are known,  $x_1,x_2$ and  $x_3$, the general solution can be found from 
the cross ratio relation \cite{CarRamcinc}
$$
\frac{x-x_1}{x-x_2}:\frac{x_3-x_1}{x_3-x_2}=k\ ,
$$
which provides us a nonlinear superposition rule.

All these properties can be better understood in the framework of Lie systems \cite{LS},
conveniently extended in some cases to include second-order differential
equations systems. These systems have a lot of applications not only in mathematics but also in
many different branches of physics, both at the classical and the quantum
level \cite{Win83}, some of them determined by 
$t$-dependent Hamiltonian that in general 
are not easy to handle. Furthermore, 
Lie systems are also relevant in Control Theory.

The paper is organized as follows. Next section is devoted to first recall some
important properties of Lie systems,  and then  to extend the theory for systems of
second-order differential equations.  Moreover, we  illustrate the theory with
some important examples with applications in physics and other relevant
examples as Milne-Pinney \cite{Mil30} equation $\ddot x=-\omega^2(t)x+
k/{x^3}$,
which  is usually studied
together with the time-dependent harmonic oscillator  $\ddot y+\omega^2(t)
y=0$, the so-called  Ermakov system.   Pinney showed   in a short
 paper \cite{P50} that the general solution of the first equation 
can be written as a 
nonlinear superposition of  two solutions of the associated
harmonic oscillator. A 
generalisation of Ermakov systems and  the corresponding Ermakov invariants,
which  appear as first integrals
in a natural way, will also be analysed. It will be shown that all these 
properties can be better understood in the framework of Lie systems \cite{LS},
conveniently extended in some cases to include second-order differential
equations systems.
The perspective of such systems as  Lie systems is very important 
because they have a lot of applications not only in mathematics but also in
many different branches of physics, both at the classical and the quantum
level \cite{Win83}. In particular, some  $t$-dependent Hamiltonian  systems
are of this class, and 
Lie systems are also relevant in Control Theory.

Section 3 is devoted to explain the reduction method which is the main ingredient in
the setting of sufficient integrability conditions that are proposed in Section
4. This is based on the affine action of the  group of curves in the associated
Lie group on the set of Lie systems with such a group (see \cite{CRPraga} for a
geometric interpretation) and generalizes the results obtained in \cite{CarRam}
for Riccati equation. The integrability conditions are used in Section 5 for
studying Caldirola--Kanai oscillator \cite{Cal41,{Ka48}} and the possibility of reduction to an
autonomous system is proved as a consequence of the theory. Arbitrary
time-dependent systems are also studied and explicit time-dependences are
proved to satisfy the above mentioned integrability conditions and therefore to 
correspond to explicitly solvable models.

\section{Lie systems: A geometric approach}

The conditions for  the system (\ref{nonasys}) to admit a
superposition rule, i.e. there exists a function 
$\Phi:{\mathbb{R}}^{n(m+1)}\to {\mathbb{R}}^n$,
$x=\Phi(x_{(1)}, \ldots,x_{(m)};k_1,\ldots,k_n)$, such that its general
solution can be written as 
$x(t)=\Phi(x_{(1)}(t), \ldots,x_{(m)}(t);k_1,\ldots,k_n)$, where 
$\{x_{(a)}(t)\mid a=1,\ldots,m\}$ is a fundamental set of particular solutions
and  $k=(k_1,\ldots,k_n)$ is a set of   $n$ arbitrary  constants, were  studied
by Lie \cite{LS}. Under very general conditions \cite{CGM07}  the necessary and sufficient
conditions is that the associated $t$-dependent vector field $X(x,t)$
can be written as a linear combination
\begin{equation}
X(x,t)=\sum_{\alpha =1}^r b_\alpha(t)\, X_{(\alpha )}(x)\ ,
\label{Lievf}
\end{equation}
such that the vector fields 
 $\{X_{(\alpha)}\mid \alpha=1,\ldots,r\}$ generate a
 finite-dimensional  real
 Lie algebra, i.e. there exist  
 $r^3$ real numbers $c_{\alpha\beta}\,^\gamma$, such that
\begin{equation}
[X_{(\alpha)},X_{(\beta)}]=\sum_{\gamma=1}^r c_{\alpha\beta}\,^\gamma
X_{(\gamma)}\ ,\qquad \alpha,\beta=1,\ldots,r\,.
\end{equation}
Linear systems are particular examples whose associated Lie algebra is
$\goth{gl}(n,\mathbb{R})$ and $m=n$ in the homogeneous case, or the
 corresponding affine
algebra and $m=n+1$ in the inhomogeneous one. Riccati equation is another example for which 
$X(x,t)=b_0(t)Y_0(x)+b_1(t)Y_1(x)+b_2(t)Y_2(x)$ with
\begin{equation} 
Y_0(x)=\pd{}x\,,\qquad Y_1(x)=x\,\pd{}x\,,\qquad Y_2(x)=x^2\,\pd{}x\,.\label{vfRicc}
\end{equation} 
with commutation relations
$$
[Y_0,Y_1]=Y_0,\quad [Y_2,Y_1]=-Y_2,\quad [Y_2,Y_0]=-2Y_1,
$$
closing on a $\goth{sl}(2,\mathbb{R})$ Lie algebra
\cite{CarRam,{CLR07b}}. There is 
an action $\Phi_{\rm Ric}:SL(2,\mathbb{R})\times \bar \mathbb{R}\rightarrow\bar
\mathbb{R}$ of the Lie group $SL(2,\mathbb{R})$ on  $\bar\mathbb{R}\equiv \mathbb{R}\cup \infty$ given by
$$
\begin{array}{rcl}
\Phi_{\rm Ric}(A,x)&=&{\dfrac{\alpha x+\beta}{\gamma x+\delta}}\ ,\ \ \ 
\mbox{if\ }\  x\neq-{\dfrac{\delta}{\gamma}}\ ,\cr
\Phi_{\rm Ric}(A,\infty)&=&{\alpha}/{\gamma}\ ,\ \ \ \ \Phi_{\rm
  Ric}(A,-{\delta}/{\gamma})=\infty\ ,,\end{array}
$$
where the matrix $A\in  {SL(2,\mathbb{R})}$ is 
$$
A=\matriz{cc} {{\alpha}&{\beta}\\{\gamma}&{\delta}}\,,
$$
such that the 
fundamental vector fields of this action are the linear combinations of the vector fields
$Y_0, Y_1$ and $Y_2$.

Another very relevant example of  Lie system is given by a  $t$-dependent
right-invariant vector field in a Lie group $G$.  A  right-invariant vector field $X^{\tt R}$ is one such that 
$X^{\tt R}(g)=R_{g*e}X^{\tt R}(e)$.
If   $\{{\rm a}_1,\ldots,{\rm a}_r\}$ is 
a basis of $T_eG$ and define the right-invariant vector fields $X^{\tt
  R}_\alpha$ by $X^{\tt R}_\alpha(g)=R_{g*e}{\rm a}_\alpha$, then the 
 $t$-dependent right-invariant vector field 
$\bar X(g,t)=-\sum_{\alpha=1}^r b_\alpha(t)X^{\tt R}_\alpha(g)$
defines a Lie system in $G$ whose integral curves 
are  solutions of the system
$\dot g=-\sum_{\alpha=1}^r b_\alpha(t)\,X^{\tt R}_\alpha(g)$,
and when applying $R_{g^{-1}}$ to both sides we see that $g(t)$ satisfies
\begin{equation}\label{eqLG}
R_{g^{-1}(t)*g(t)}\dot g(t)\,=-\sum_{\alpha=1}^r b_\alpha(t){\rm a}_\alpha\in T_eG
\ .\label{lieingr}
\end{equation}
Right-invariance means that it is enough to know one solution, for instance the
one starting from the neutral element $g(t)$, to know all the solutions of the
equation with any initial condition, i.e. we obtain the solution $g'(t)$  with
initial condition $g'(0)=g_0$ as $R_{g(0)}g(t)$. 
A generalization of the method by
Wei-Norman \cite{WN} is very useful in solving such equation, and furthermore reduction
techniques can also be used \cite{CarRamGra}. Finally, as right-invariant
vector fields $X^{\tt R}$ project on the fundamental vector fields in each homogeneous
space for $G$, the solution of (\ref{lieingr}) allows us to find the general
solution for the corresponding Lie system in each homogeneous
space. Conversely, the knowledge of particular solutions of the associated 
system in a homogeneous space gives us a method for reduction the problem to
the corresponding isotopy group \cite{CarRamGra}. This equation is also 
important
 because any Lie system described by a $t$-dependent vector field in a manifold
 $M$ like (\ref{Lievf}) where the vector fields are complete
  and  close the same
commutation
 relations as the basis $\{{\rm a}_1,\ldots,{\rm a}_r\}$ determines an action
$\Phi_{\rm LieG}:G\times M\rightarrow M$ such that the  vector field
$X_\alpha$ is the
fundamental 
vector field corresponding to  ${\rm a}_\alpha$, and moreover, the integral
curves of the t-dependent vector field are obtained from the solutions of  thr  equation (\ref{eqLG}). More explicitly,
 the general solution of the given  Lie system are
 $x(t)=\Phi_{\rm LieG}(g(t),x_0)$, where $x_0$ is
 an initial condition of the solution and $g(t)$ is the solution for (\ref{eqLG}) with $g(0)=e$.

The search for the number $m$ of solutions and the superposition function
$\Phi$ have recently been studied from a geometric perspective
\cite{CGM07}. Essentially we should consider  `diagonal prolongations' to
$\mathbb{R}^{n(m+1)}$,  $\widetilde
X(x_{(0)}, \ldots, x_{(m)},t)$, of the $t$-dependent  vector field
$X(x,t)=X^i(t,x)\,\partial/\partial x^i$, given by
\[
\widetilde X(x_{(0)}, \ldots,x_{(m)},t)=\sum_{a=0}^mX_a(x_{(a)},t)\,,\qquad
t\in {\mathbb{R}}\,,
\]
where $X_{a}(x_{(a)},t)=X^i(x_{(a)},t)\,{\partial}/{\partial x^i_{(a)}}$,
such that the extended system admits $n$ independent integrals of motion, which
define in an implicit way the superposition function.

As another example of Lie system of physical relevance,
consider the $t$-dependent harmonic oscillator  described by a
Hamiltonian 
$$
H(x,p,t)=\frac 12\,\frac {p^2}{m(t)}+\frac 12\, m(t)\omega^2(t)x^2\,,
$$
whose  Hamilton equations  are given by
$$
\left\{\begin{array}{rl}
\dot x&=\dfrac{\partial H}{\partial p}=\dfrac p{m(t)}\\
\dot p&=-\dfrac{\partial H}{\partial x}=-m(t)\omega^2(t)\, x\,.
\end{array}\right.
$$
Its
solutions are the integral curves of the $t$-dependent vector field 
\begin{equation}
X(x,p,t)=\frac 1{ m(t)}\,p\,\pd{}x-m(t)\omega^2(t)\,x\pd{}p\,.\label{vfTDHO}
\end{equation}
If we consider the set of vector fields
$$
X_0=p\pd{}{x},\quad X_1=\frac 12\left(x\pd{}{x}-p\pd{}{p}\right),\quad X_2=-x\pd{}{p}\,,
$$
which close on a $\goth{sl}(2,\mathbb{R})$ Lie algebra with commutation relations
$$
[X_0,X_1]=X_0,\quad [X_2,X_1]=-X_2,\quad [X_2,X_0]=-2X_1,
$$
then   the $t$-dependent vector field (\ref{vfTDHO}) 
can be written as a linear combination
 with $t$-dependent coefficients 
$
X(\cdot,t)=\sum_{\alpha=0}^2b_\alpha(t)X_\alpha(\cdot)$,
 with 
 $$b_0(t)=\frac 1{m(t)}\,, \quad b_1(t)=0\,,\quad b_2(t)=m(t)\omega^2(t)\,.$$
Remark that the vector fields of this example and those of Riccati equation 
 close the same commutation relations.

As an instance of t-dependent harmonic oscillator we next develop the one with damping term, with equation of
motion 
$$\frac d{dt}(m_0\, \dot x)+ m_0\,\mu\,\dot x+k\, x=0\ ,\qquad k=m_0\omega^2\,,
$$
admits a Hamiltonian description with a $t$-dependent Hamiltonian \cite{Cal41,{Ka48}}
\begin{equation}
H=\frac{p^2}{2m_0}\exp(-\mu t)+\frac{1}{2}m_0\exp(\mu t)\omega^2x^2=\frac
{p^2}{2m(t)}+
\frac{1}{2}m(t)\omega^2x^2\label {HTDHO}
\end{equation}
with $m(t)=m_0\,\exp(\mu t)$. The  time-dependent coefficients as a Lie system are: 
$$b_0(t)=\frac{\exp(-\mu t)}{m_0}\,, \quad b_1(t)=0\,,\quad  b_2(t)=m_0\,\exp(\mu t)\,\omega^2.
$$

We can also consider the time-dependent frequency harmonic oscillator
\begin{equation}
H=\frac 12p^2+\frac 12F(t)\omega_0^2x^2\,,\label{HTDFHO}
\end{equation}
whose associated  $t$-dependent vector field $X$ is
 a linear combination
$ X(\cdot,t)=F(t)\omega_0^2X_2(\cdot)+X_0(\cdot)$ with coefficients: 
$$b_0(t)=1\,, \quad b_1(t)=0\,,\quad b_2(t)=F(t)\omega_0^2\,.
$$

The theory of Lie systems can be extended to include systems of second-order
differential equations.  Such a system 
 whose  associated system of first-order differential equations is a Lie system is
  called SODE Lie system \cite{CLR07}. This is the
case of the 1-dimensional  harmonic oscillator (\ref{howt}), where $X(x,v,t)$ is a linear
combination $X(\cdot,t)= X_2(\cdot)-\omega^2(t)X_1(\cdot)$ 
of the vector fields 
$X_1(x,v)= x\,{\partial}/{\partial v}$ and $ X_2(x,v)=v\,{\partial}/{\partial x}$ that close a finite-dimensional Lie algebra with $X_3(x,v)=1/2(x\,{\partial}/{\partial x}-v\,{\partial}/{\partial v})$. 
Similarly for 
the isotropic 2-dimensional case,
\begin{equation}
\left\{\begin{array}{rcl}\ddot x_1&=& -\omega^2(t) x_1\cr\ddot x_2&=& -\omega^2(t)
  x_2\end{array}\right.\label{2dimho}
\end{equation}
but with
$X_1(x_1,x_2,v_1,v_2)= x_1\,{\partial}/{\partial v_1}+ x_2\,{\partial}/{\partial v_2}$, 
$X_2(x_1,x_2,v_1,v_2)=v_1 \,{\partial}/{\partial x_1}+v_2 \,{\partial}/{\partial x_2}$ and
$X_3(x_1,x_2,v_1,v_2)=1/2(x_1 \,{\partial}/{\partial x_1}+x_2 \,{\partial}/{\partial x_2}-v_1 \,{\partial}/{\partial v_1}-v_2 \,{\partial}/{\partial v_2})$,
in both cases  with associated Lie algebra $\goth{sl}(2,\mathbb{R})$,

The general theory of Lie systems can now be used and so we can obtain 
 a first integral, 
$F(x_1,x_2,v_1,v_2)=x_1v_2-x_2v_1$, which is nothing but the Wronskian of two solutions, or the
angular momentum from the physical viewpoint. This may  be seen as a partial superposition rule: if $x_1(t)$ is a
solution of the first equation, then we obtain for each real number $k$
 the first-order differential
equation for the variable $x_2$, 
$x_1(t)\, {dx_2}/{dt} =k+\dot x_1(t)x_2$,
from where $x_2$ can be found to be given by  
$$
x_2(t)=k' x_1(t)+k\, x_1(t)\int^t\frac{d\zeta}{x_1^2(\zeta)}\,.
$$
In a similar way, 
with three copies of the same harmonic oscillator, one obtains two first
integrals,  $F_1(x_1,x_2,x,v_1,v_2,v)=xv_1-x_1v$ and
$F_2(x_1,x_2,x,v_1,v_2,v)=xv_2-x_2v$ that 
 produce a superposition
rule, because from
$$\left\{\begin{array}{crl} xv_2-x_2v&=&k_1\cr x_1v-v_1x&=&k_2\end{array}\right.
$$ 
 we obtain the expected superposition rule \cite{CLR07}:
$$x=k_1\, x_1+k_2\, x_2\,,\qquad v=k_1\, v_1+k_2\, v_2\,.
$$

It has only recently been pointed out \cite{CLR07} that the differential
equation \cite{Mil30,{P50}}
\begin{equation}
\ddot x=-\omega^2(t)x+\frac k{x^3}\,,\label{Milneeq}
\end{equation}
is also a Lie system
with associated Lie algebra $\goth{sl}(2,\mathbb{R})$:  the corresponding
$t$-dependent vector field is
\begin{equation}
X(x,v,t)=v\pd{}x+\left(-\omega^2(t)x+\frac k{x^3}\right)\pd{}v\,.\label{Milnevf}
\end{equation}
which is a linear combination $X=L_2-\omega^2(t)L_1$ with 
$$
L_1=x\frac {\partial}{\partial v},\qquad L_2=\frac k {x^3}\frac{\partial}{\partial v}
+v\frac{\partial}{\partial x}, 
$$
which are such that 
$$
[L_1,L_2]=2L_3,\quad [L_3,L_1]=L_1,\quad [L_3,L_2]=-L_2
$$
with 
$$
 L_3=\frac 1 2 \left(x\frac{\partial}{\partial x}-v\frac{\partial}{\partial v}\right)\,.
$$

Another important example of Lie system with the same Lie algebra
$\goth{sl}(2,\mathbb{R})$ is the   Ermakov system:
\begin{equation}\label{Ermak}
\left\{\begin{array}{rcl}
\dot x&=&v_x\cr
\dot v_x&=&-\omega^2(t)x\cr
\dot y&=&v_y\cr
\dot v_y&=&-\omega^2(t)y+\dfrac 1 {y^3}\nonumber
\end{array}\right.
\end{equation}
with associated $t$-dependent vector field 
$$X=v_x\pd{}x+v_y\pd{}y-\omega^2(t)x\pd{}{v_x}+\left(-\omega^2(t)y+\frac 1
  {y^3}\right)\pd{}{v_y}\,,
$$
which is a linear combination with $t$-dependent coefficients, $X=-\omega^2(t)X_1+X_2$, with
$$X_1=x\pd{}{v_x}+y\pd{}{v_y}\,,\qquad X_2=v_x\pd{}x+v_y\pd{}y+\frac 1
  {y^3}\pd{}{v_y}\,.
$$ 
This system is made up  by two Lie systems  closing on
 a $\goth{sl}(2,\mathbb{R})$ algebra: the first one is a harmonic
 oscillator  and the  second one is 
a Pinney equation. The system admits a
 first integral of the  motion which can be found as a solution of 
$X_1F=X_2F=0$ because these two conditions imply that $X_3F=1/2[X_1,X_2]F=0$. But $X_1F=0$ 
means that $F(x,y,v_x,v_y)=\bar F(x,y,\xi)$ with $\xi=xv_y-yv_x$,
and then $X_2F=0$ is written
$$
v_x\pd{\bar F}{x}+v_y\pd{\bar F}{y}+\frac x{y^3}\pd{\bar F}{\xi}=0\,,
$$ and  from the associated characteristics system 
we obtain $$\frac {x\,dy-y\, dx}{\xi}=\frac {y^3\,d\xi}{x}\Longrightarrow
  \frac{d(x/y)}{\xi}+\frac{y\,d\xi}{x}=0
$$
 the following first integral is found:
$$
\psi(x,y,v_x,v_y)=\left(\frac{x}{y}\right)^2+\xi^2=\left(\frac{x}{y}\right)^2+(xv_y-yv_x)^2
$$
which is the well-known Ermakov invariant. 
       A possible generalisation of the                     Ermakov system (\ref{Ermak})
   is  given by:
\begin{equation}
\left\{\begin{array}{rcl}
\ddot{x}&=&{\displaystyle\frac{1}{x^3}}f(y/x)-\omega^2(t)x\cr
\ddot{y}&=&{\displaystyle\frac{1}{y^3}}g(y/x)-\omega^2(t)y
\end{array}\right.,
\end{equation}
which reduces to (\ref{Ermak}) for 
                       $f(u)=0$ and $g(u)=1$.

This system can be written as  a first-order one by doubling the number of
degrees of freedom by introducing the new variables   $v_x$ and $v_y$:
$$
\left\{\begin{array}{rcl}
\dot x&=&v_x\\
\dot v_x&=&-\omega^2(t)x+\frac 1 {x^3} f( y/x)\\
\dot y&=&v_y\\
\dot v_y&=&-\omega^2(t)y+\frac 1 {y^3} g(y/x)
\end{array}\right.
$$
which determines the integral curves of the vector field 
$$X=v_x\,\pd{}{x}+v_y\,\pd{}{v_y}+\left(-\omega^2(t)x+\frac 1 {x^3} f( y/
  x)\right)\pd{}{v_x}+\left(-\omega^2(t)y+\frac 1 {y^3} g(y/
    x)\right)\pd{}{v_y}\,.
$$
Such vector field can be written as a linear combination 
$$X=N_2-\omega^2(t)\, N_1
$$
where $N_1$ and $N_2$ are the vector fields 
$$
N_1=x\frac{\partial }{\partial v_x}+y\frac{\partial }{\partial v_y},\quad N_2=v_x\frac{\partial}{\partial x}+
\frac{1}{x^3}f(y/x)\frac{\partial}{\partial v_x}+v_y\frac{\partial}{\partial y}+
\frac{1}{y^3}g( y/x)\frac{\partial}{\partial v_y},
$$

Note that these vector fields generate a three-dimensional real Lie algebra
with a third generator $$N_3=\frac 12\left(x\frac{\partial}{\partial
    x}-v_x\frac{\partial}{\partial v_x}+y\frac{\partial}{\partial
    y}-v_y\frac{\partial}{\partial v_y}\right)\,.$$ 
In fact, as  
$$
[N_1,N_2]=2N_3, \quad [N_3,N_1]=N_1, \quad  [N_3,N_2]=-N_2
$$
they generate a Lie algebra isomorphic to   $\goth{sl}(2,\mathbb{R})$. Therefore the
system is a Lie system.

There exists  a first  integral for the motion, $F:\mathbb{R}^4\rightarrow \mathbb{R}$,
for any $\omega^2(t)$,  which can be found in a similar way and we will arrive
to the first integral:
$$
\frac 12 (xv_y-yv_x)  ^2+\int^{x/y}\left[-\frac 1{u^3}\, f\left(\frac 1u\right)+
  u\,g\left(\frac 1u
\right)\right]\,du\,.
$$

 This
first  integral allows us  to determine  a solution of one subsystem 
 in terms of  a solution of the other equation.

Finally, we can revisit 
Pinney equation  by
considering a system 
 made up by a
  Pinney equation in the $x$ variable with two associated harmonic oscillator
 equations for variables $y$ and $z$.
by 
considering  the system of first-order differential equations:
$$
\left\{
\begin{array}{rcl}
\dot x&=&v_x\cr
\dot y&=&v_y\cr
\dot z&=&v_z\cr
\dot v_x&=&-\omega^2(t)x+
\dfrac k {x^3} \cr
\dot v_y&=&-\omega^2(t)y\cr
\dot v_z&=&-\omega^2(t)z
\end{array}\right.
$$
which corresponds to the vector field 
$$X=v_x\pd{}x+v_y\pd{}y+v_z\pd{}z+\frac{k}{x^3}
\frac{\partial}{\partial v_x}-\omega^2(t)\left(x\frac{\partial }{\partial v_x}+y\frac{\partial }{\partial v_y}+
z\pd{}{v_z}\right)
$$
The vector field $X$ can be expressed as $X=N_2-\omega^2(t)N_1$ where the vector
fields $N_1$ and $N_2$ are:
$$
N_1=x\frac{\partial }{\partial v_x}+y\frac{\partial }{\partial v_y}+
z\pd{}{v_z},\quad N_2=v_x\frac{\partial}{\partial x}+
\frac{1}{x^3}\frac{\partial}{\partial v_x}+v_y\frac{\partial}{\partial
  y}+v_z\frac{\partial}{\partial z},
$$ 
These vector fields generate a     3-dimensional real Lie algebra with 
                $N_3$ given by 
$$
 N_3=\frac 12\left(x\frac{\partial}{\partial x}-v_x\frac{\partial}{\partial
     v_x}+y\frac{\partial}{\partial y}-v_y\frac{\partial}{\partial
     v_y}+z\frac{\partial}{\partial z}-v_z\frac{\partial}{\partial
     v_z}\right)\,.
$$
In fact, as  
$$
[N_1,N_2]=2N_3, \quad [N_3,N_1]=N_1, \quad  [N_3,N_2]=-N_2
$$
they generate a Lie algebra isomorphic to   $\goth{sl}(2,\mathbb{R})$. Thus this
       is a Lie system.

There exist  three
first-integrals for  the distribution generated by these fundamental vector
fields:
 the Ermakov invariant
 $I_1$ of the subsystem involving variables $x$ and $y$, 
the Ermakov invariant $I_2$ 
of the subsystem involving variables $x$ and $z$, and finally,  
the Wronskian $W$ of the subsystem involving variables $y$ and $z$.
They are given by $W=yv_{z}-zv_{y}$,
$$I_1=\frac 12\left((yv_{x}-xv_y)^2+k\left(\frac {y}x\right)^2\right)\,,\quad
I_2=\frac 12\left((xv_{z}-zv_x)^2+k\left(\frac {z}x\right)^2\right)
\,.
$$

In terms of these three integrals we can obtain an explicit expression of $x$ in terms of $y, z$ and the integrals $I_1, I_2, W$:
$$
x=\frac {\sqrt{2}}{W}\left(I_2y^2+I_1z^2\pm\sqrt{4I_1I_2-kW^2}\ yz
\right)^{1/2}\,.
$$

We can recover  in this way the result of  \cite{P50}: the general
  solution of Pinney equation can be expressed  in terms of two 
solution of the corresponding harmonic oscillator problem
with time-dependent frequency

\section{The reduction method}

Given an equation (\ref{lieingr}) 
 on a Lie group, it may happen that the only non-vanishing coefficients
are those corresponding to a subalgebra $\goth h$ of $\goth g$ and then the equation reduces to a simpler equation on a subgroup, involving less coordinates.
An important result is that if we know a particular solution of the
 problem associated in a homogeneous space, the original solution reduces
 to one on the isotopy subgroup \cite{CarRamGra}.

One can show that there is an action of the group 
 $\mathcal{G}$ of curves in $G$ on the set of right-invariant Lie systems in
 $G$ 
(see e.g. \cite{CRPraga} for a geometric justification), 
and we can take advantage of such
 an action for transforming a given Lie system into another simpler one. 

So, if $g(t)$ is a solution of the given Lie system satisfying (\ref{eqLG}) and 
we choose a curve $g^\prime(t)$ in the group $G$, 
 and define 
a curve $\overline g(t)$ by $\overline g(t)=g^\prime(t)g(t)$, then 
  the new curve in 
$G$, $\overline g(t)$, determines a
 new Lie system. 
Indeed,
$$
R_{\overline g(t)^{-1}* \overline g(t)}(\dot {\overline g}(t))
=R_{g^{\prime\,-1}(t)*g^{\prime}(t)}(\dot g^\prime(t))
-\sum_{\alpha=1}^r b_\alpha(t){\rm Ad\,}(g^{\prime}(t)){\rm a}_\alpha\ , 
$$
which is similar to the original one, 
with a different right-hand side. 
Therefore, the aim is
to choose the curve $g^\prime(t)$  in such a way that 
 the new equation be simpler. 
For instance, we can choose a subgroup $H$ and 
look for a choice of $g'(t)$ such that the right hand side 
 lies in $T_e H$, and hence if $\overline g(0)=e$ then $\overline g(t)\in H$ for all
$t$. This can be done when we know  a 
 solution of the associated Lie system 
in $G/H$, what allows us to reduce the problem to one in the subgroup $H$, see \cite{CarRamGra}.  
\begin{theorem}
Each solution of (\ref{lieingr})
on the group $G$ can be written in the 
form $g(t)=g_1(t)\,h(t)$,
where $g_1(t)$ is a curve on $G$ projecting onto a solution 
$\tilde g_1(t)$  for the left action $\lambda$ of $G$ on
the homogeneous space $G/H$
and $h(t)$ is a solution of an equation but for the
subgroup $H$,
given explicitly by
$$
(R_{h^{-1}*h}\dot h\, )(t) 
=-{\rm Ad\,}(g_1^{-1}(t))\left(\sum_{\alpha=1}^r b_\alpha(t){\rm a}_\alpha
+(R_{g_1^{-1}*g_1}\dot g_1)(t)\right)\in {T_eH}\ .
$$
\end{theorem}

This fact is very important because one can show that
Lie systems associated 
with solvable Lie algebras are solvable by quadratures
 and therefore,
given a Lie system with an arbitrary $G$ having a solvable subgroup, we should
look for a possible transformation from the original system to one which reduces
to the subalgebra and therefore integrable by quadratures. 

The result of the preceding  Theorem proves that for a solvable Lie subgroup $H$ of $G$
there always exists a curve in $G$ that transforms the initial Lie system into
 a new one in subgroup  $H$ of $G$. Nevertheless, it can be difficult to find out a solution of the equation in $G/H$ that determines this transformation. Then, to be able
to obtain one is more interesting to suppose also that this transformation is a
curve in a certain subset of $G$, i.e. a one-dimensional Lie subgroup. 
When  such a transformation 
 exists, it  is  easier
 to obtain it, but it may be that such a transformation does not exist, and 
  {\sl 
the conditions for the existence of such a transformation of a certain form are integrability
conditions for the system}.

We could choose for showing this assertion a particular 
example: Riccati equation.
One can find in the literature a lot of integrability criteria for Riccati
equation, all of them particular examples of the above method.

We can also consider more relevant examples in Physics, for instance,
time-dependent harmonic oscillators.

The results obtained for one system are valid for the other; they are
essentially conditions for the equation in the group, and 
both are examples of Lie systems associated with the same Lie group:
$SL(2,\mathbb{R})$.

\section{Integrability criteria for Lie systems}

Consider the particular case of Lie systems with associated Lie group 
$SL(2,\mathbb{R})$, therefore valid for Riccati equation (\ref{ricceq}),
Milne-Pinney 
equation (\ref{Milneeq}) 
and the time-dependent
harmonic oscillator described by (\ref{vfTDHO}). 
The group $SL(2,\mathbb{R})$ contains the affine group (either the one generated by
$X_0$ and $X_1$ or the one generated by $X_1$ and $X_2$), which is solvable. Therefore, a
transformation from the given equation to one of this subgroup allows us to express the
general solution in terms of quadratures.
This happens when we know a particular solution $x_1$ of the given equation:
$x=x_1+z$, what corresponds to choose 
$${\bar g}(t)=\matriz{cc}{1&-x_1\cr 0&1}\,,
$$
reduces the equation  to
$$
\frac {dz}{dt}=(2\, a_2\, x_1+a_1) z+a_2\,z^2
$$
The reduction by the knowledge of two or three quadratures has also been
studied from this perspective and similarly Strelchenya criterion \cite{{CarRam},{Stre}}.

Each Riccati equation can be considered as a 
curve in ${\mathbb{R}}^3$ and   we can 
transform every function in $\mathbb{R}$, $x(t)$,
under an element of the group ${\GR}$
of smooth $SL(2, \mathbb{R})$-valued curves 
${\rm Map}(\mathbb{R},\,SL(2,\mathbb{R}))$, as follows:
$$
\begin{array}{rcl}
\Theta(A,x(t))&=&{\dfrac{\alpha(t) x(t)+\beta(t)}{\gamma(t) x(t)+\delta(t)}}\ ,\ \ \ 
\mbox{if\ }\  x(t)\neq-{\dfrac{\delta(t)}{\gamma(t)}}\ ,\cr
\Theta(A,\infty)&=&{\alpha(t)}/{\gamma(t)}\ ,\ \ \ \ \Theta(A,-{\delta(t)}/{\gamma(t)})=\infty\ ,\cr &&\mbox{when}\ 
A=\matriz{cc} {{\alpha(t)}&{\beta(t)}\\{\gamma(t)}&{\delta(t)}}\,\in{\cal G}\
.\end{array}
$$
The image $x'(t)=\Theta({\bar A}(t),x(t))$ of 
 a curve $x(t)$ solution of the given Riccati equation
satisfies a new Riccati equation with coefficients $b'_2,b'_1, b'_0$, given by 
$$
\begin{array}{rcl}
b'_2&=&{\bar\delta}^2\,b_2-\bar\delta\bar\gamma\,b_1+{\bar\gamma}^2\,b_0+\bar\gamma{\dot{\bar\delta}}-\bar\delta \dot{\bar\gamma}\ ,
\nonumber \cr
b'_1&=&-2\,\bar\beta\bar\delta\,b_2+(\bar\alpha\bar\delta+\bar\beta\bar\gamma)\,b_1-2\,\bar\alpha\bar\gamma\,b_0   
       +\bar\delta\dot{\bar\alpha}-\bar\alpha \dot{\bar\delta}+\bar\beta \dot{\bar\gamma}-\bar\gamma \dot{\bar\beta}\ ,   \cr
b'_0&=&{\bar\beta}^2\,b_2-\bar\alpha\bar\beta\,b_1+{\bar\alpha}^2\,b_0+\bar\alpha\dot{\bar\beta}-\bar\beta\dot{\bar\alpha} \ .
 \nonumber 
\end{array}
$$
This expression defines an affine action of the group  ${\GR}$ on the set of 
Riccati equations. 
As indicated above  the vector fields defining  the TDHO and the Riccati
equation  close the same conmutation relations and both examples 
are related with exactly the same equation in $SL(2,\mathbb{R})$, but they
correspond to different actions.

 Lie systems in $SL(2,\mathbb{R}) $ defined by a constant curve,
${\rm a}(t)= {\displaystyle \sum_{\alpha=0}^2} c_\alpha {\rm a}_\alpha$,
are integrable and  the same happens for 
curves of the form ${\rm a}(t)= D(t)\left({\displaystyle \sum_{\alpha=0}^2} c_\alpha {\rm
    a}_\alpha\right)$, where $D$ is an arbitrary function, because a time parametrisation reduces the problem to
the previous one.

The system is essentially a Lie system on a one-dimensional Lie group.

A straightforward application of the reduction method leads to
 the following result which is valid not only for
 Riccati equation but also for any other Lie system with Lie group
 $SL(2,\mathbb{R})$ in which the 
$t$-dependent vector field be described in terms of a set of vector fields
closing on  the same commutation relations as those of  Riccati equation. In
this way, all these systems 
with the same $t$-dependent coefficients are related with the same equation in
$SL(2,\mathbb{R})$
 and this allows us to generalise  for all these systems the integrability
 condition obtained in the next theorem. 
The other results can be generalised by means of changing the action of the
Riccati equation for the action 
of the new  Lie system.

\begin{theorem}
The necessary and sufficient conditions
for the existence of a  transformation:
$$
y'=G(t)y \qquad {\it  with }\qquad  G(t)\geq 0 
$$
generated by the transformation
$$
y'=\Phi_{\rm Ric}(\bar A(t),y) \qquad {\it  with  }\qquad  \bar A(t)=\matriz{cc}{\alpha(t)&0\\0&\alpha^{-1}(t)}
$$
and relating the Riccati equation in an interval $I\in \mathbb{R}$
$$
\frac{dy}{dt}=b_0(t)+b_1(t)y+b_2(t)y^2\,,  \qquad \forall t\in I,\quad (b_0b_2)(t)\ne 0, 
$$with an integrable one given by
$$
\frac{dy'}{dt}=D(t)(c_0+c_1y'+c_2y'^2)\,,
$$
where $c_i$ are real numbers,  $c_i\in \mathbb{R}$, is that for such constants 
 $c_0c_2\ne 0$ and 
$$
D^2(t)c_0c_2=b_0(t)b_2(t),\qquad \frac{b_1(t)+\frac{1}{2}\left(\frac{\dot b_2(t)}{b_2(t)}-\frac{\dot b_0(t)}{b_0(t)}\right)}{D(t)}=c_1.
$$
The  transformation is then uniquely defined by:
$$
y'=\sqrt{\frac{b_2(t)c_0}{b_0(t)c_2}}\,y\,.
$$
\end{theorem}
As a consequence of this theorem, given the above Riccati equation
if there are constants $K$ and $L$ such that 
$$
 \left(b_1(t)+\frac{1}{2}\left(\frac{\dot b_2(t)}{b_2(t)}-\frac{\dot b_0(t)}{b_0(t)}\right)\right)\sqrt{\frac{L}{b_0(t)b_2(t)}}=K\,,
$$
then there exists a 
 time-dependent linear change of variables  transforming the given equation into
 the solvable  Riccati equation:
$$
\frac{dy'}{dt}=D(t)(c_0+c_1y'+c_2y'^2) \label{ricc1dim}
$$
where
$c_1=K$, ${c_0c_2}=L$,
and $D(t)$ is given as above. 
The existence of such constants $K$  and $L$ can be considered a sufficient condition for
integrability of the given Riccati equation. Indeed, $L$ is just used to make the square to exists and can always be obtained, the real integrability condition is the existence of the constant $K$.

\section{Some applications in physics}

\subsection{The Caldirola--Kanai              oscillator}
Coming back to the time-dependent harmonic oscillator, we see that
the Caldirola--Kanai model is an example of application of the previous Theorem.

We have seen that it is a Lie system with group $SL(2,\mathbb{R})$ and associated
  coefficients 
$$b_0(t)=\frac 1{m_0}\,\exp(-\mu t)\,, \quad b_1(t)=0\,,\quad  b_2(t)=m_0\,\omega_0^2\exp(\mu t)\,.
$$
Therefore, as $b_0(t)b_2(t)=\omega_0^2$, $b_1=0$ and 
$({\dot b_2}/{b_2}-{\dot b_0}/{b_0}=2\mu$, 
we see that the integrability condition of the preceding theorem holds with $K=\mu/\omega_0$ and $L=1$ , and the function $D$ is
then a constant $D=\omega_0$.  Therefore we choose $c_0=1, c_1=\mu/\omega_0$ and $c_2=1$ and this
example reduces to the system
$$\frac{d}{dt}\matriz{c}{x\\p}=\matriz{cc}{\mu&\omega_0\\-\omega_0&-\mu}\matriz{c}{x\\p}$$
which can be easily integrated. This shows once again that the Caldirola--Kanai
model
can be reduced to an autonomous system, as a particular case of a more general situation.
\subsection{The $t$-dependent frequency harmonic oscillator}
In the case of the $t$-dependent frequency harmonic oscillator for which
 $\omega^2(t)=F(t)\omega_0$,
$ X(\cdot,t)=F(t)\omega_0^2X_2(\cdot)+X_0(\cdot)$,
i.e. $b_0(t)=1$, $b_1(t)=0$ and $b_2(t)=F(t)\omega_0^2$, 
the condition on $F$ to satisfy the compatibility condition is
$$
\frac 12\, \frac{\dot F} F=K\,\omega_0\, \sqrt{F}
$$
and therefore $F$ must be of the form
$$
F(t)=\frac{1}{(-K\omega_0 t+K')^2}\,,
$$
for another constant $K'$ and then the Hamiltonian which can be exactly integrated is
$$
H=\frac{p^2}{2}+\frac{1}{2}\frac{\omega_0^2}{(-K\omega_0 t+K')^2}x^2
$$
and the corresponding Hamilton equations are
$$
\left\{\begin{array}{rl}
\dot x&=p\\
\dot p&=-\dfrac{\omega_0^2}{(-K\omega_0 t+K')^2}\,x
\end{array}\right.
$$

Had we used instead the family of curves in $SL(2,\mathbb{R})$ given by
$$
\bar A_0(t)=\left(\begin{array}{cc}\frac{1}{V(t)}&0\\-u_1& V(t)\end{array}\right)
$$
where $u_1$ is a constant, and we would obtain the following relations among coefficients:
$$
\left\{\begin{array}{rl}
b'_2&= V^2b_2+u_1Vb_1+u_1^2b_0-u_1\dot V\\
b'_1&=b_1+2\dfrac {u_1}V-2\dfrac{\dot V}V\\
b'_0&=\dfrac 1{V^2}b_0\end{array}\right.
$$
Assume   we want to relate the $t$-dependent vector field
$X(\cdot ,t)=X_0+F(t)\omega_0^2X_2$,
characterised by
$b_0=1,\quad b_1=0,\quad b_2=F(t)\omega_0^2$,
with another one characterised by $b'_0,b'_1$ and $b'_2$, which is  integrable,
or more explicitly, related with the $t$-dependent vector field 
$X(\cdot,t)=D(t)(c_0X_0+c_2X_2)$, 
i.e. $b'_0=Dc_0$, $b'_1=0$, and $b'_2=Dc_2$,

Then $b_1=b'_1=0$, and the second  equation shows that $\dot V=u_1$,
i.e. $V(t)=u_1t+u_0$ with $u_0\in \mathbb{R}$, and using this condition in the
first equation, together with $b_0=1$,  it becomes $b'_2= V^2b_2$,
and then, as the third equation gives us the value of $D$ as $c_0\,D=b'_0=1/V^2$,
we see that $b'_2=Dc_2=V^2F(t)\omega_0^2$, and therefore $F$ must be 
 proportional to $(u_1t+u_0)^{-4}$:
$$F(t)=\frac k{(u_1t+u_0)^{4}}\,, \qquad k=\frac{c_2}{c_0\omega_0^2}\,.
$$
We can fix $c_2=\omega_0^2$ and $c_0=1$. Thus
$F(t)=(u_1t+u_0)^{-4}$.

Finally, note that the time-dependent transformation $\bar A_0(t)$ which performs the reduction is
$$
\left\{\begin{array}{rl}
x'&=\dfrac{x}{V(t)}\\
p'&=-u_1x+V(t)p
\end{array}\right.
$$
 which transforms  the initial system of  differential equations into
$$
\left\{\begin{array}{rl}
\dfrac{dx'}{dt}&=\dfrac{1}{V^2(t)}p'\\
\dfrac{dp'}{dt}&=\dfrac{1}{V^2(t)}(-\omega_0^2x')
\end{array}\right.
$$
Now, using  the time-reparametrization
$$
\tau(t)=\int^t_0\frac{dt'}{V^2(t')}
$$
we obtain the following time-independent linear system
$$
\left\{\begin{array}{rl}
\dfrac{dx'}{d\tau}&=p'\\
\dfrac{dp'}{d\tau}&=-\omega_0^2x'
\end{array}\right.
$$
whose  general solution is
$$
\left(\begin{array}{c}x'(\tau)\\ p'(\tau)\end{array}\right)=
\left(\begin{array}{cc}\cos(\omega_0 \tau)&\frac{\sin(\omega_0 \tau)}{\omega_0}\\-\omega_0\sin(\omega_0 \tau)&\cos(\omega_0 \tau)\end{array}\right)
\left(\begin{array}{c}x'(0)\\ p'(0)\end{array}\right)
$$
and thus we obtain that
$$
x(t)=V(t)\left(\cos(\omega_0\,
  \tau(t))\frac{x_0}{V(0)}+\frac{1}{\omega_0}\sin(\omega_0\,\tau(t))(-u_1x_0+V(0)p_0)\right)
$$
where we have used 
$$
x'(0)=\dfrac{x_0}{V(0)}\,,\qquad 
p'(0)=-u_1 x_0+V(0)p_0\,.$$

The same computations are valid for 
the Pinney equation when $F(t)\omega_0^2$ is replaced by a time-dependent angular 
frequency $\omega^2(t)$.

\section* { Acknowledgments}
 Partial financial support by research projects MTM2006-10531 and E24/1 (DGA)
and a F.P.U. grant from  Ministerio de Educaci\'on y Ciencia are acknowledged.


\section*{References}
\begin{thebibliography}{10}
\bibitem{CarRam}
J.F. Cari\~nena and  A. Ramos,
{Int. J. Mod. Phys.}  {\bf A 14}, (1999) 1935--51.
\bibitem
{LS}  S. Lie,
{\sl Vorlesungen \"uber continuierliche Gruppen mit 
Geometrischen und anderen Anwendungen\/},
Edited and revised by G. Scheffers,
Teubner, Leipzig, 1893.
\bibitem{Win83}
{P. Winternitz},
{\it Lie groups and solutions of nonlinear differential equations},
 in: {\sl Nonlinear Phenomena}, K.B. Wolf Ed., Lecture Notes in Physics {\bf 189},
{Springer-Verlag, N.Y., 1983}
\bibitem{CGM00}  
J.F. Cari\~nena,  J. Grabowski and G. Marmo,
{\sl Lie--Scheffers systems: a geometric approach\/},
Bibliopolis, Napoli, 2000.
\bibitem{CGM07}  
J.F. Cari\~nena,  J. Grabowski and G. Marmo,
Rep. Math. Phys. {\bf 60}, (2007) 237--58
\bibitem{AHW81} R.L.  Anderson, J. Harnad and P. Winternitz, 
Lett. in Math. Phys. {\bf  5},  (1981)  143-148.
\bibitem{HarWinAnd83}
J. Harnad, P. Winternitz and R.L. Anderson,   
J. Math. Phys. {\bf 24},  (1983) 1062--72.
\bibitem{SW84} S. Shnider  and  P. Winternitz,
J. Math. Phys. {\bf 25}, (1984) 3155--65. 
\bibitem{OlmRodWin86a}
M.A. del Olmo, M.A. Rodr\'{\i}guez and P. Winternitz, 
{J. Math. Phys.} {\bf 27}, (1986) 14--23 
\bibitem{OlmRodWin86b}
M.A. del Olmo, M.A. Rodr\'{\i}guez and P. Winternitz, 
{\it Integrability, chaos and nonlinear superposition formulas for differential
matrix Riccati equations}, in: {\sl Quantum Chaos and Statistical Nuclear
Physics}, pp. 372-378, 
Lecture Notes in Physics {\bf 263}, 1986, 
 \bibitem{OlmRodWin87}
M.A. del Olmo, M.A. Rodr\'{\i}guez and P. Winternitz, 
J. Math. Phys. {\bf 28},  (1987) 530--535.
\bibitem{CarRamcinc}
{J.F. Cari\~nena and  A. Ramos},
Acta Appl. Math. {\bf 70},   (2002) 43--69.
\bibitem{CLR07b}
J.F. Cari\~nena, A. Ramos and J. de Lucas, 
Elect. J. Diff. Eqs. {\bf  122}, (2007) 1--14.
\bibitem{Mil30}  W.E. Milne, 
Phys. Rev. {\bf 35}, (1930) 863--67.
\bibitem{P50} E. Pinney,  
 Proc. A.M.S. {\bf 1},  (1950) 681.
\bibitem{CRPraga}
J.~F. Cari\~nena  and  A. Ramos, {\it Lie systems and Connections in fibre
bundles: Applications  in  Quantum Mechanics},
 in: {\sl 9$^{\rm th}$ Int. Conf. Diff. Geom and Appl.}, p. 437--452    (2004), J.
Bures {\sl et al.}  Eds.,   (Matfyzpress, Praga,  2005).
\bibitem{Cal41}P. Caldirola, 
Nuovo Cim. {\bf  18}, (1941) 393--400. 
\bibitem{Ka48}E. Kanai, 
Prog. Theor. Phys. {\bf 3}, (1948) 440--42.
\bibitem{WN}  {J. Wei and E. Norman},
 {J. Math. Phys. \bf 4},  (1963) 575--81.
\bibitem{CarRamGra}  
J.F. Cari\~nena, J. Grabowski and  A. Ramos,
 Acta Appl.  Math. {\bf 66}, (2001) 67--87.
\bibitem{CLR07}  J.F. Cari\~nena, J. de Lucas  and M.F.  Ra\~nada,
{\it Nonlinear superpositions and Ermakov systems}, in: {\sl
  Differential Geometric Methods in Mechanics and Field Theory}, pp. 15--33,
eds F. Cantrijn, M. Crampin and B. Langerock (Academia Press, 2007)
\bibitem{Stre} V. M. Strelchenya,
J. Phys. A: Math. Gen. {\bf 24},  (1991) 4965--4967.

\end{thebibliography}
 \end{document}